\documentclass[twocolumn,aps,prl,amsmath,amssymb]{revtex4}
\usepackage{graphicx}
\usepackage{dcolumn}
\usepackage{bm}
\usepackage[dvips]{color}
\usepackage{epstopdf}
\DeclareGraphicsExtensions{.pdf,.eps,.png,.jpg,.mps}
\begin{document}
\preprint{AIP/123-QED}

\title{Surface plasmon polariton graphene photodetectors}
\author{T. J. Echtermeyer$^{1,2}$}
\author{S. Milana$^1$}
\author{U. Sassi$^1$}
\author{A. Eiden$^1$}
\author{M. Wu$^1$}
\author{E. Lidorikis$^3$}
\author{A. C. Ferrari$^1$}
\email{acf26@eng.cam.ac.uk}
\affiliation{$^1$Cambridge Graphene Centre, University of Cambridge, Cambridge CB3 0FA, UK}
\affiliation{$^2$School of Electrical \& Electronic Engineering, University of Manchester, Manchester M13 9PL, UK}
\affiliation{$^3$Department of Materials Science and Engineering, University of Ioannina, Ioannina 45110, Greece}
\begin{abstract}
The combination of plasmonic nanoparticles and graphene enhances the responsivity and spectral selectivity of graphene-based photodetectors. However, the small area of the metal-graphene junction, where the induced electron-hole pairs separate, limits the photoactive region to sub-micron length scales. Here, we couple graphene with a plasmonic grating and exploit the resulting surface plasmon polaritons to deliver the collected photons to the junction region of a metal-graphene-metal photodetector. This results into a 400\% enhancement of responsivity and a 1000\% increase in photoactive length, combined with tunable spectral selectivity. The interference between surface plasmon polaritons and the incident wave introduces new functionalities, such as light flux attraction or repulsion from the contact edges, enabling the tailored design of the photodetector's spectral response. This architecture can also be used for surface plasmon bio-sensing with direct-electric-readout, eliminating the need of complicated optics.
\end{abstract}
\maketitle
Graphene-based photodetectors (PDs)\cite{Ferrari2014Nanoscale, Koppens2014NatNano} have been reported with ultra-fast operating speeds (up to 262GHz from the measured intrinsic response time of graphene carriers\cite{Urich2011}) and  broadband operation from the visible and infrared\cite{Xia2009,Lee2008,Mueller2010,Urich2011,Mueller2009,Lemme2011,Echtermeyer2011,Liu2011_2,Gabor2011,Konstantatos2012,Engel2011,Yan2012,Chitara2011,Freitag2013} up to the THz\cite{Vicarelli2012NatMat, Spirito2014APL, Cai2014NatNano}. The simplest graphene-based photodetection scheme relies on the metal-graphene-metal (MGM) architecture\cite{Lee2008,Mueller2009,Park2009,Lemme2011,Gabor2011,Song2011,EchtermeyerPolPap}, where the photoresponse is due to a combination of photo-thermoelectric and photovoltaic effects\cite{Lee2008,Mueller2009,Park2009,Lemme2011,Gabor2011,Song2011,EchtermeyerPolPap}. For both mechanisms, the presence of a junction is required to spatially separate excited electron-hole (e-h) pairs\cite{Lee2008,Mueller2009,Park2009,Lemme2011,Gabor2011,Song2011,EchtermeyerPolPap}. At the metal-graphene junction, a work-function difference causes charge transfer and a shift of the graphene Fermi level underneath the contact\cite{Lee2008,Xia2009,Mueller2009,Giovanetti2008}, compared to that of graphene away from the contact\cite{Lee2008,Xia2009,Mueller2009,Giovanetti2008}, resulting into a build-up of an internal electric field (photovoltaic mechanism)\cite{Lee2008, Mueller2009,Peters2010APL,Rao2011ACSNano} and into a difference of Seebeck coefficients (photo-thermoelectric mechanism)\cite{Gabor2011,Song2011, Sun2012NatNano}. An alternative way to create a junction is to use a set of gate electrodes to electrostatically dope graphene\cite{Lemme2011,Gabor2011}.

For both photovoltaic and photo-thermoelectric mechanisms, however, the spatial extend of the junction is$\sim$100-200nm\cite{Mueller2009,Lemme2011,Gabor2011}, which reduces the photoactive area to a fraction of the diffraction limited laser spot size in a typical scanning current microscopy experiment. Furthermore, suspended undoped graphene only absorbs 2.3\% of light\cite{Nair2008} which, while remarkably high for a one atom thick material, is low in absolute terms for practical applications. This is further reduced by a factor of $4/(1+n)^2$ for graphene on a dielectric substrate of refractive index $n$ (see Methods). Additionally, in highly doped graphene the absorption decreases even further\cite{Mak2008,Lagatsky2013}.

One approach is to extend the junction region in order to capture more light. In a vertical (i.e. with doping profile perpendicular to the device's surface) p-i-n semiconductor PD this is achieved by ion-implantation with tailored dose and energy\cite{Sze1981}. In the MGM configuration, however, the lateral nature (i.e. doping profile parallel to the device's surface) of the junction does not allow a straightforward doping profile engineering and, thus far, to the best of our knowledge, no techniques have been reported to reliably do that. As an alternative, several graphene-based vertical architectures have been proposed, including all-graphene\cite{Kim2014NatComm}, or graphene integrated with semiconductor layers, such as other two dimensional (2d) materials\cite{britnell2013Scince,Roy2013NatNano} or Si\cite{An2013NL,Liu2014ACSNano}. In the latter cases, however, graphene is not the absorbing material and the spectral response is thus far limited to above the band gap of the semiconductor layer. Furthermore, while these approaches have demonstrated high responsivities (up to$\sim5\times10^8$A/W in Ref.\cite{Roy2013NatNano} by employing MoS$_2$ as light capturing material) they do come with the cost of smaller operation speed (up to 100kHz in Ref.\cite{Kim2014NatComm}) as compared to the graphene-based PDs operating at speeds up to 50 GBit/s in the optical link reported in Ref.\cite{Schall2014}.

Improving graphene absorption in the ultra-fast MGM configuration is thus critical. Various solutions have been proposed, such as the integration of graphene into an optical microcavity\cite{Furchi2012NL,Engel2012NatComm} ($>$20-fold enhancement) or onto a planar photonic crystal cavity\cite{Shiue2013APL} (8-fold enhancement), to take advantage of the multiple passes of the trapped light through graphene, or its coplanar integration with a Si integrated photonic waveguide\cite{Pospischil2013NatPhot,Gan2013NatPhot,Wang2013NatPhot} ($>$10-fold enhancement). Another solution is the integration of plasmonic nanostructures on graphene\cite{Echtermeyer2011,Liu2011_2,Fang2012} to exploit the strongly enhanced electromagnetic near-fields\cite{Schedin2010,Maier2007} associated with the localized surface plasmon resonances (LSPR)\cite{Schedin2010,Maier2007}. LSPRs originate from the resonant coherent oscillation of the metal's conduction electrons in response to the incident radiation. The resulting enhanced near-fields surrounding the nanostructures promote light absorption in the materials around them\cite{Schedin2010}. We previously reported a $\times20$ enhancement in photoresponse\cite{Echtermeyer2011} when radiation is focused close to the nanostructures. In this approach, however, light absorbed around nanostructures far from the junction (where efficient charge separation occurs) does not fully contribute to the photoresponse\cite{Fang2012}.

An ideal alternative would be to enable light collection in one part of a device and then guide it into the junction region at the contact edge. This can be achieved by exciting surface plasmon polaritons (SPPs) on the metal contacts. SPPs are surface-bound waves propagating on a metal-dielectric interface and originate from the coupling of light with the metal's free electrons\cite{Maier2007,raether,economou69}. Their excitation can be achieved by means of an integrated diffraction grating\cite{Maier2007,raether}, and their delivery to the active region (junction at contact edge) will enhance the overall absorption. Thus, the contact now becomes a light collector. Such an approach was demonstrated with semiconductor-based near infra red (NIR) PDs\cite{Ishi2005,Ren2011NL,Yu2006,Bhat2008,Shackleford2009,Karar2011APL} in order to reduce the semiconductor active area without compromising light absorption. A smaller active area results into reduced carrier transit time\cite{Ishi2005,Ren2011NL} and reduced capacitance\cite{Ishi2005,Ren2011NL}, thus increased operation speed. In particular, a circular (bull's eye) grating\cite{Ishi2005,Ren2011NL} was used to deliver SPPs into a subwavelength circular aperture on top of vertical Si\cite{Ishi2005} or Ge\cite{Ren2011NL} Schottky photodiodes, while a linear grating was used to deliver SPPs into a subwavelength linear slit in a lateral metal-GaAs-metal photodiode configuration\cite{Shackleford2009,Karar2011APL}. Operation speeds beyond 100GHz where estimated\cite{Ishi2005,Ren2011NL}, while responsivity enhancements (compared to a device without the grating) were up to $\times4$ for linear gratings\cite{Shackleford2009,Karar2011APL} and over $\times10$ for circular gratings\cite{Ishi2005,Ren2011NL}. No compromise in operation speed was reported due to the presence of the grating\cite{Shackleford2009}. This approach was also applied in mid-IR detection\cite{Harrer2014APL}, by delivering light into a quantum cascade detector, and in THz detection\cite{Wang2012APL}, by delivering light into a GaAs/AlGaAs 2d-electron-gas bolometer.

Here we apply the SPP grating coupler concept to a MGM PD and demonstrate a$\sim$400\% increase in responsivity and a$\sim$1000\% increase in photoactive length. Furthermore, we show that this offers a solution to another problem: in order to have a net response under uniform illumination of the whole MGM PD area, one must break the mirror symmetry between the two contacts\cite{Mueller2010}. In contrast to the metal-semiconductor-metal case, applying a bias is not practical because it would result into a large dark current, due to the semimetallic nature of graphene\cite{Mueller2010}. Using different metallizations for the two contacts is an option\cite{Mueller2010}, but increases the fabrication steps. In our approach this problem is addressed by using different contact grating structures. One can utilize the interference between SPPs and incident waves and create novel asymmetric contact designs that produce complex spectral responses, such as switching the light flux between the two contacts edges, enabling new functionalities, such as label- and optics-free direct-electrical-readout plasmonic biosensing.

To explore the design opportunities offered by a SPP grating coupler on the metal contacts, we first carry out numerical simulations using the finite-difference time-domain (FDTD) method\cite{taflove,Schedin2010}. The SPP wavevector on a metal-dielectric interface is\cite{Maier2007}:
\begin{equation}
k_{SPP}=(\omega/c)\sqrt{(\epsilon_m \epsilon_d)/(\epsilon_m + \epsilon_d)}
\label{eq:spp}
\end{equation}
where $\epsilon_m(\omega)$, $\epsilon_d(\omega)$ are the dielectric functions for the metal and dielectric medium respectively, and the SPP existence condition is $\epsilon_m<-\epsilon_d$\cite{Maier2007,raether}. $k_{SPP}$ is larger than any propagating wave in the dielectric, whose wavevector is $k=(\omega/c)\sqrt{\epsilon_d}$. This momentum mismatch between SPPs and propagating waves implies that SPPs cannot decay into free propagating waves, but also that one cannot directly excite SPPs from free waves on a smooth metal surface\cite{Maier2007}. One way to overcome this is by diffraction, whereby the continuity of the component of momentum parallel to the surface is broken and incident light can scatter into SPPs\cite{Maier2007}. This can be achieved by a nano-slit\cite{tejeira07NatPhys,choi09APL}, a diffracting element\cite{radko08PRB,sondergaard12PRB}, or a grating coupler\cite{radko08PRB,leveque06JAP}. In the latter case, in particular, a periodic array of metal ridges and grooves delivers the additional momentum according to\cite{Maier2007}:
\begin{equation}
k_{SPP} = k_{||}\pm mK
\label{eq:spp2}
\end{equation}
where $k_{||}=(\omega/c)\sin{\theta}$ is the parallel component of incident wavevector, $\theta$ the incident angle, $K=2\pi/a$ the grating's reciprocal lattice vector, $a$ the grating pitch, and $m$ the diffraction order.

\begin{figure}
\centerline{\includegraphics[width=90mm]{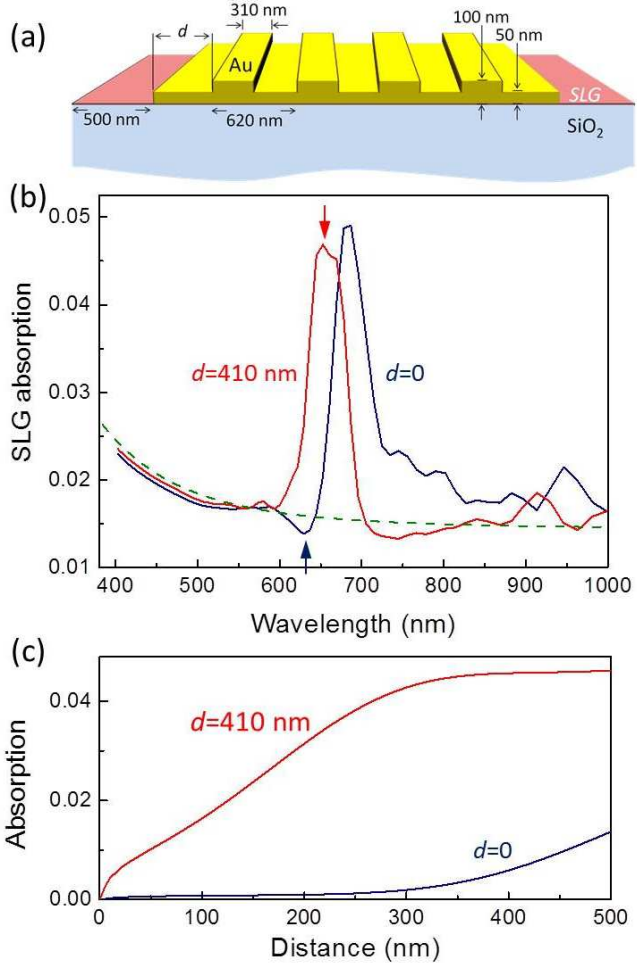}}
\caption{(a) Schematic of the simulated structure. (b) Relative (i.e. normalized to the case without the grating) absorption in the exposed SLG. (c) Cumulative absorption in the exposed SLG integrated away from the contact edge, for the two cases pointed by the arrows.}
\label{fig:theory1}
\end{figure}

We select a grating of 50nm Au bars periodically placed at a pitch of 620nm on top of a 50nm Au contact film, as depicted in Fig.\ref{fig:theory1}a, at a 1:1 ratio of ridge and groove widths. These grating parameters are chosen since they were shown to be optimal for yielding a high percentage of incident light scattered into SPPs ($\sim$20\% in Refs.\cite{radko08PRB,leveque06JAP}). A termination "step" of width $d$ extends beyond the last ridge. For simplicity, we assume the graphene/contact structure to be on top of a semi-infinite SiO$_2$ substrate. The dielectric functions of Au\cite{johnsonchristy2} and graphene\cite{kravets10} are treated through a Drude-Lorentz model, as explained in Ref.\cite{Schedin2010} and in Methods. Inserting Au's dielectric function in Eqs.\ref{eq:spp},\ref{eq:spp2}, with $\epsilon_d$=1 for air, yields $\lambda_0$=645nm for the vacuum wavelength of the SPPs on the Au/air interface. It is also possible to have SPPs in the Au/SiO$_2$ interface. Given that $\epsilon_d$=2.13 for SiO$_2$\cite{palik}, they are excited at $\lambda_0$=930nm.

In the first set of calculations the device is illuminated by a normally incident plane wave polarized perpendicular to the grating (transverse magnetic-TM). This polarization is required because the SPPs are themselves TM waves\cite{Maier2007,raether}, having both longitudinal (parallel to the propagation direction) and transverse (perpendicular to the surface) electric field components. The exposed part of the single layer graphene (SLG) in-between the contacts is kept fixed at 1000nm, while the width of the Au contact, thus of the SLG unexposed (buried) part, is varied depending on the number of ridges and the size $d$ of the termination step. Periodic and absorbing boundaries are assumed in the lateral and vertical directions respectively (see Methods). Fig.\ref{fig:theory1}b plots the absorption in the exposed SLG normalized to the flux incident on the exposed SLG area, for $d$=0 and 410nm. Strong enhancement peaks, reaching up to 5\% absorption, are found between 630-700nm. There are also secondary small peaks at$\sim$900-950nm, assigned to the Au/SiO$_2$ interface SPPs. The green dashed line indicates the absorption within the 1000nm wide SLG on top of the semi-infinite SiO$_2$ substrate in the absence of the grating, as derived in the thin film limit [see Methods]:
\begin{equation}
A_{SLG}=4A_{SLG}^{0}/(n_{SiO_2}+1)^2
\label{eq:theory1}
\end{equation}
where $A_{SLG}^{0}=\pi e^2/\hbar c=2.3\%$ is the absorption coefficient for suspended SLG in air\cite{Nair2008} and $n_{SiO_2}$=1.46 is the SiO$_2$ refractive index in the above wavelength range\cite{palik}. A three-fold wavelength-selective increase in absorption is observed, due to coupling with SPPs scattered from the grating.

We now consider the parameter $d$. Fig.\ref{fig:theory1}b indicates that for certain wavelengths (e.g.$\sim$630nm), there are opposite extremes of absorption for the two different steps. Such an asymmetry can be instrumental in designing contact layouts that exhibit a net photovoltage even under uniform illumination. For the two cases pointed by the two arrows of Fig.\ref{fig:theory1}b, we report in graph Fig.\ref{fig:theory1}c the cumulative absorption in the exposed SLG as we move away from the grating edge. In the $d$=410nm case, absorption is strongly enhanced at the edge of the grating, and starts leveling off$\sim$300nm away from it. In contrast, for $d$=0 absorption is minimal close to the grating and starts picking up only$>$300nm from the grating edge. SPP interference causes one contact to "pull" light close to its edge and the other to "push" it away from it. The situation reverses at$\sim$700nm. To confirm that these light "attraction" and "expulsion" effects are not related to interferences within the contact, i.e. that they are independent of contact width, we perform the same calculations for 9, 11 and 13 ridges. While we observe some small interference effects at longer wavelengths ($>$730nm, especially for $d$=0), within the primary wavelength range of interest (630-700nm) the absorption is independent of grating size. Furthermore, the response is well saturated above 10 grating periods, consistent with numerical studies on the influence of the number of grating periods\cite{Yu2006,Bhat2008,Ren2011NL}.
\begin{figure}
\centerline{\includegraphics[width=80mm]{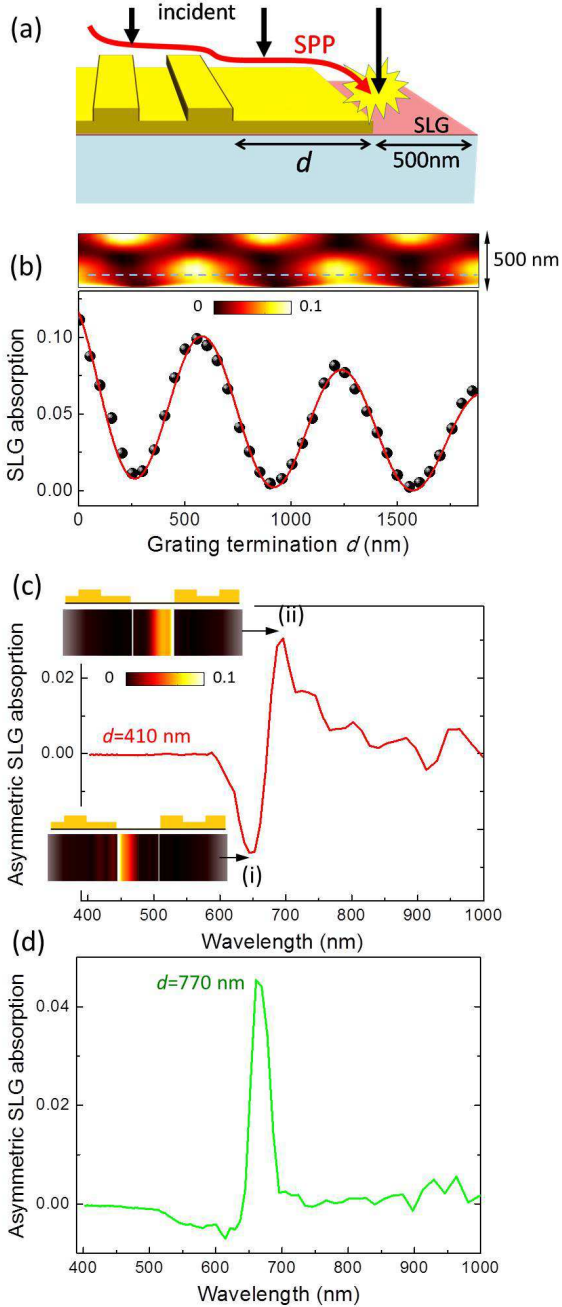}}
\caption{(a) Depiction of the  interference mechanism. (b) Relative (i.e. normalized to the case without the grating) SLG absorption within 100nm from the contact edge as a function of $d$. The line is a fit of Eq.\ref{eq:spp_interf}, which assumes SPP+incident wave interference. The top inset color-codes the absorption distribution in the exposed SLG as a function of $d$. (c) Asymmetric contact layout with $d$=0 and 410nm, as shown in the insets, with 600nm exposed SLG width. The asymmetric absorption (i.e. $A_R-A_L$, where $A_R$, $A_L$ are the SLG absorptions in the right and left halves of the SLG channel), is proportional to the net photovoltage under coherent uniform illumination. The insets color-code the absorption profiles for the two peaks indicated by the arrows. (d) Asymmetric absorption when the 410nm contact is replaced by a 770nm one.}
\label{fig:theory1b}
\end{figure}

The mechanism responsible for these sharp contrasts in absorption is interference: the excited SPPs travel down the termination step towards the exposed SLG and interfere with the incident waves there, as depicted in Fig.\ref{fig:theory1b}a. Interference between SPP waves was reported in pure metal systems\cite{tejeira07NatPhys,choi09APL,weeber07NL,brongersma07NatNan}. To confirm that what shown in Fig.\ref{fig:theory1}b is due to interference of SPPs with the incident wave, we examine a series of $d$ values, and plot the absorption in the exposed SLG within the first 100nm from the contact edge. These calculations are performed at 680nm, i.e. in-between the two peaks in Fig.\ref{fig:theory1}b. An oscillatory response is observed in Fig.\ref{fig:theory1b}b (points), which is characteristic of interference between two waves with a variable phase between them. The total field at the SLG will be $E_{inc}+E_{SPP}$, where $E_{inc}=1$ for an incident wave normalized to unit amplitude, and $E_{SPP}=\Delta \mathrm{e}^{i(k_{SPP} \cdot d+\phi_0)}$ is the SPP amplitude, with $\Delta$ the relative SPP electric field strength compared to the input field, $k_{SPP}$ the SPP wave vector calculated from Eq.\ref{eq:spp} for the Au/air interface, and $\phi_0$ a constant phase. We thus fit the SLG absorption of Fig.\ref{fig:theory1b}b to:
\begin{equation}
A=A_{SLG}\left|1+\Delta \mathrm{e}^{i(k_{SPP} \cdot d+\phi_0)}\right|^2
\label{eq:spp_interf}
\end{equation}
where $A_{SLG}$ is evaluated from Eq.\ref{eq:theory1}. As discussed later, SPPs on the Au/air interface can leak into the dielectric substrate\cite{drezet08MSEB} providing an extra loss mechanism. To account for these losses we scale the imaginary part of the wave vector $k_{SPP} = Re\left\{k_{SPP}\right\}+i Im\left\{k_{SPP}\right\}$ according to $Im\left\{k_{SPP}\right\}\rightarrow s\cdot Im\left\{k_{SPP}\right\}$. We treat $\Delta$, $\phi_0$ and $s$ as adjustable parameters and fit Eq.\ref{eq:spp_interf} to the simulation of Fig.\ref{fig:theory1b}a (note that the oscillation period is dependent only on $Re\left\{k_{SPP}\right\}$). An excellent fit (line) is obtained with $\Delta$=1.87, $\phi_0=\pi$/5 and $s$=9, confirming SPP excitation, propagation and interference. Fig.\ref{fig:theory1b}b (top) color codes the absorption within the first 500nm of exposed SLG as a function of $d$. The oscillation of light "attraction" and "expulsion" from the contact edge is apparent.

The implication of Eq.\ref{eq:spp_interf} is is that the asymmetry in photovoltage is larger than what seen in Fig.\ref{fig:theory1}b. E.g., if the two contacts shown in Fig.\ref{fig:theory1}b are placed across each other with a 600nm gap, then all the light accumulated from both contacts will be funneled close to only one of them. Such a scheme offers great flexibility in designing asymmetric contacts suitable for uniform illumination, potentially eliminating the need for different metallizations\cite{Mueller2010}. We explore this asymmetric contact design in Fig.\ref{fig:theory1b}c (see insets). The asymmetric absorption is defined as $A_{asym}=A_R-A_L$, where $A_{R}$ is the SLG absorption in the 300nm area close to the right ($d$=0) contact and $A_{L}$ is the SLG absorption in the 300nm area close to the left ($d$=410~nm) contact. $A_{asym}$ is then proportional to the expected net photovoltage. An antisymmetric response is obtained, as shown in Fig.\ref{fig:theory1b}c. In particular, at 650nm (point (i) in Fig.\ref{fig:theory1b}c) all the flux is "pulled" to the left contact, while at 700nm (point (ii) in Fig.\ref{fig:theory1b}c) all the flux is "pushed" to the right contact. I.e., a spectrally-selective region less than 50nm wide is created, within which the photovoltage abruptly changes sign. Outside this region ($\lambda<$580nm and $\lambda>$800nm) both contacts have similar responses, thus the net photovoltage is zero. Yet, this is not the only response function available. Plotted in Fig.\ref{fig:theory1b}d is the case of a left contact with $d$=770nm. The response is found to be symmetric around 660nm. In both cases, the peak absorption is 2-3 times higher than what a 600nm SLG on SiO$_2$ would absorb in the absence of the contacts. In addition, optimizing the contacts just for the highest absorption (i.e. without creating "clear" symmetric or antisymmetric response functions), the peak absorption exceeds 6\%, i.e. a 4-fold increase compared to SLG on SiO$_2$. If we further limit ourselves to the first 100nm from the contact edge (i.e. within the junction), then Fig.\ref{fig:theory1b}a gives an 8-fold enhancement. Further exploration of the system's response, including absorption in the unexposed (buried) SLG under the contact, different light polarizations, unpatterned contacts and SiO$_2$ (300nm)/Si substrate, for both uniform and focused illumination, are consistent with what discussed above (see Methods).
\begin{figure*}
\centerline{\includegraphics[width=175mm]{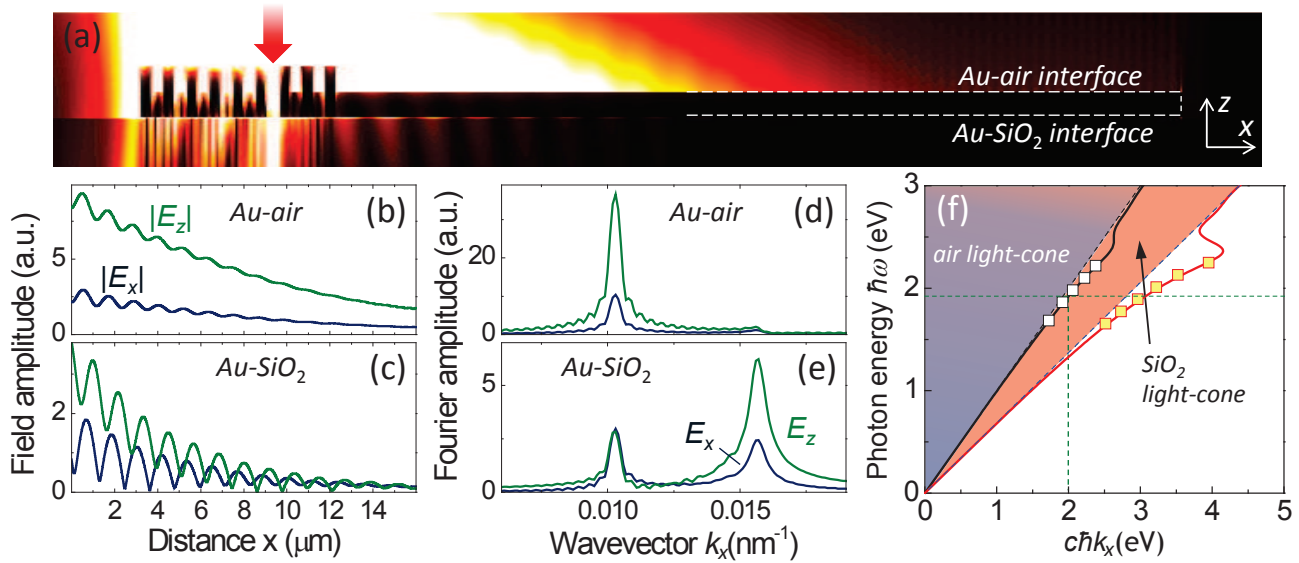}}
\caption{(a) Field intensity distribution for 650nm focused illumination (1$\mu$m spot size) of a contact with a large $d$=20$\mu$m on one side. (b,c) Field amplitude profiles along the Au-air and Au-SiO$_2$ interfaces. (d,e) Spatial Fourier amplitudes of the fields along the interfaces. (f) Dispersion relations between frequency and real part of the wavevector. The lines are from Eq.\ref{eq:spp} and the squares are the peaks of the simulated Fourier amplitudes. The vertical and horizontal dashed lines denote the SPP condition in Eq.\ref{eq:spp} with $a$=620nm. The light lines of the two dielectric media (air and SiO$_2$) extend between the vertical axis at $k_x=0$ (normal incidence) to the $k_x=n\omega/c$ line (90$^o$ incidence), defining the phase-space of allowed free-propagation modes in the index $n$ medium}
\label{fig:theory2}
\end{figure*}

A detailed analysis of all the SPPs that can be launched in our system is performed by considering a patterned contact with a large $d$=20$\mu$m, illuminated by a narrow 1$\mu$m width TM-polarized source positioned on top of the grating, as shown by the arrow in Fig.\ref{fig:theory2}a. In contrast to plane wave incidence, for which $k_{||}=0$ and where only SPPs compatible with $k_{SPP}=mK$ are excited (see Eq.\ref{eq:spp2}), the focused beam has $|k_{||}|>0$ incident wave vector components. It thus allows the full spectrum of SPPs to be excited according to Eq.\ref{eq:spp2}. Absorbing boundaries are employed in all directions to avoid scattered light from re-entering the computational cell. In Fig.\ref{fig:theory2}a the electric-field intensity distribution is plotted at 650nm. Strong scattering and fields extending many microns away from the grating are observed, both in the Au/air and Au/SiO$_2$ interfaces. An intensity oscillation is also observed in the latter, with a period$\sim$1.4$\mu$m. SPPs have both longitudinal and transverse field components, the latter being perpendicular to the metal surface\cite{Maier2007,raether}. Fig.\ref{fig:theory2}b,c plots the longitudinal $|E_x|$ and transverse $|E_z|$ electric field amplitudes at the Au/air and Au/SiO$_2$ interfaces, as a function of distance $x$ from the grating. A simple decay curve is obtained at the top interface, but an oscillating decay curve is obtained for the bottom interface. Fig.\ref{fig:theory2}d,e report the spatial Fourier transform on these fields. A single peak is found in the top interface, and two in the bottom one: one at exactly the same wavevector as in the top interface, and the other at a larger wavevector. By repeating this procedure at different illumination frequencies we get the SPP dispersion shown in Fig.\ref{fig:theory2}f. Lines denote the theoretical dispersion curves from Eq.\ref{eq:spp} using Au's dielectric function and assuming either an Au/air or Au/SiO$_2$ interface. Squares indicate the peaks obtained from the simulations by the spatial Fourier transforms. In the Au/air interface only the Au/air SPP dispersion curve emerges, while both Au/air and Au/SiO$_2$ SPP dispersions emerge in the Au/SiO$_2$ interface. SPPs are surface waves bound on a metal/dielectric interface because they exist below the light-cone of the dielectric (i.e. the phase-space of free propagating modes defined by $k_{||}\leq n\omega/c$)\cite{Maier2007,raether}. The SPP at the Au/SiO$_2$ interface is below both the air and SiO$_2$ light-cones (i.e. $k_{SPP}\geq n_{SiO_2}\omega/c$, $n_{air}\omega/c$), thus cannot couple to any free radiation states. On the other hand, the SPPs at the Au/air interface are below the air light-cone and within the SiO$_2$ light-cone (i.e. $n_{air}\omega/c\leq k_{SPP}\leq n_{SiO_2}\omega/c$), thus they can leak (tunnel) into the free radiation states of substrate, explaining why we obtain two SPP signals in the Au/SiO$_2$ interface. Such leaky waves have been used for SPP characterization within the context of Leakage Radiation Microscopy\cite{drezet08MSEB} (a far-field optical method analyzing the leaked SPP waves in glass substrates to characterize SPP propagation on the top interface of a flat or nanostructured metal film\cite{drezet08MSEB}). In a semi-infinite substrate they will just propagate away. In a finite one, on the other hand, part of the leaked waves will be reflected back to the interface and contribute more to the SLG absorption. The dominant effect in photoresponse, however, remains in the Au/air SPP, as inferred by the Fourier amplitudes in Figs.\ref{fig:theory2}b-e, and confirmed by the absorption in Fig.\ref{fig:theory1}b, with minimal contribution from Au/SiO$_2$ SPPs at 930nm.
\begin{figure}
\centerline{\includegraphics[width=90mm]{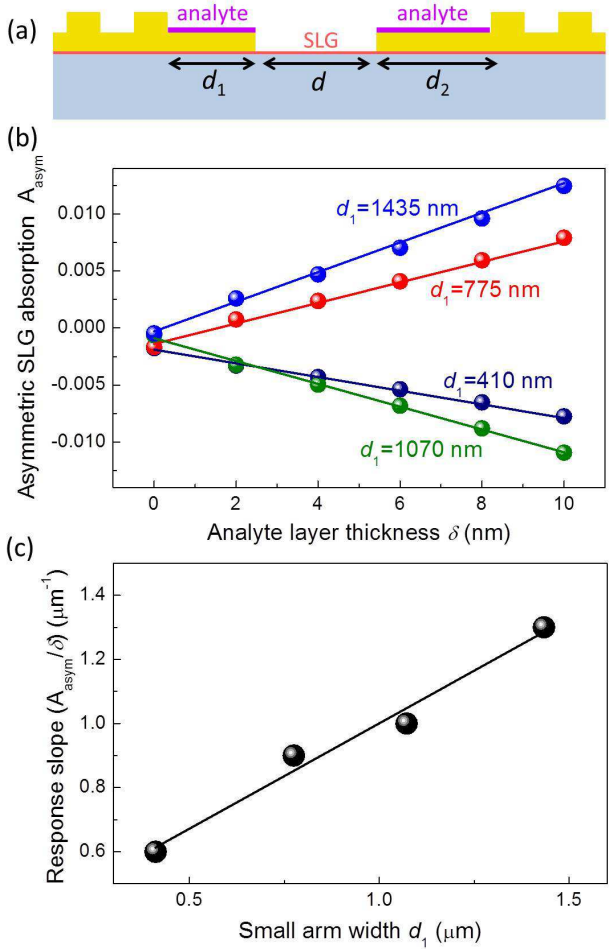}}
\caption{a) Schematic biosensing device. $d$=600nm is assumed for the exposed SLG. An analyte (purple) is deposited on both sensor arms. (b) Asymmetric absorption (proportional to the net photovoltage) as a function of analyte thickness for four sensor arm pairs. The operating wavelength is 678nm. (c) Response curve slope as a function of small sensor arm length.}
\label{fig:biosensor}
\end{figure}
\begin{figure}
\centerline{\includegraphics[width=90mm]{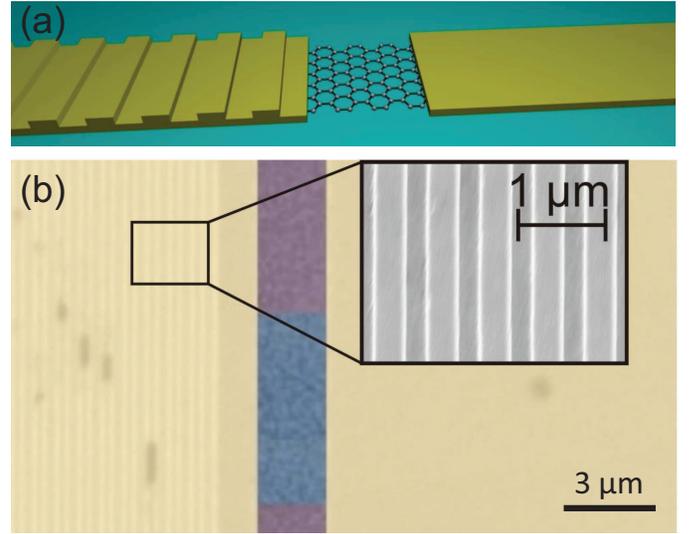}}
\caption{MGM-PD with plasmonic grating coupler. a) Schematic, b) SEM micrograph.}
\label{graph:graphene_grating}
\end{figure}

Besides photodetection, SPP+incident wave interference in a MGM architecture also lends itself to label-free surface plasmon biosensing\cite{brolo2012NatPhot, homola2008CR, gao2011ACSNano, feng2012NL}, whereby SLG assumes the role of an integrated transducer providing direct electrical readout, thus eliminating the need for optical measurements. The use of SLG as an integrated transducer was reported in a dielectric waveguide sensor geometry\cite{kim2012ABB}, but not in surface plasmon sensing. Fig.\ref{fig:biosensor} assumes that one termination step (sensor arm hereafter) has length $d_1$ and the other $d_2$, so that they are at the highest slopes of Fig.\ref{fig:theory1b}a, i.e. in the midplane of the interference oscillation, with one of them at a positive slope and the other at a negative one. In this setup, the two contacts are at an accidental degeneracy, producing the same interference between SPP and incident wave, thus zero net photovoltage under uniform illumination. If now the dielectric environment around the sensor arms changes by the presence of an analyte, it will cause an increase in $k_{SPP}$, thus an additional phase to both SPPs. Having the two arms on a different slope in the response curve of Fig.\ref{fig:theory1b}a introduces an asymmetry, thus a net photovoltage. The larger the dielectric change, the larger the photovoltage. Also, the longer the sensor arms, the higher the sensitivity, as the SPP will travel a longer distance, therefore sampling more analyte.
\begin{figure*}
\centerline{\includegraphics[width=160mm]{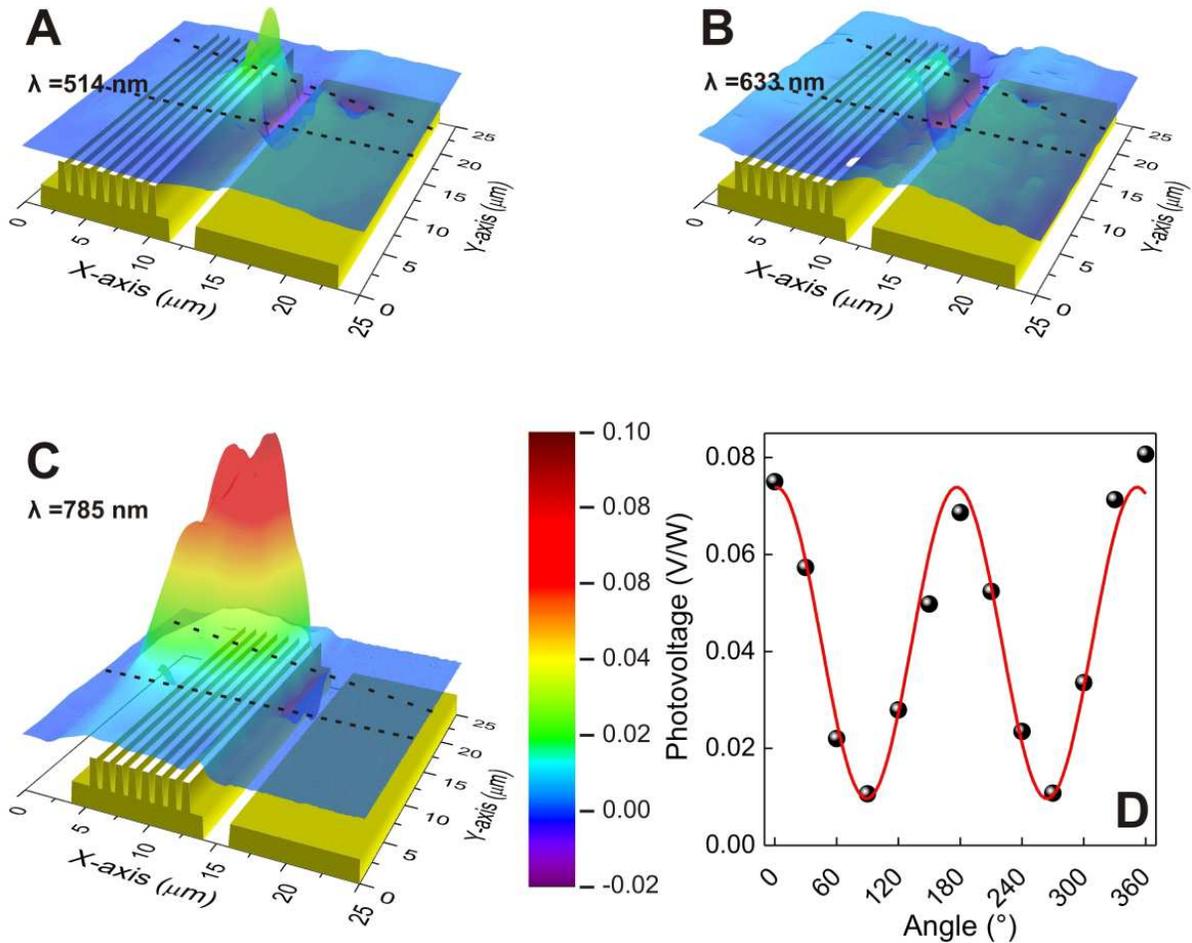}}
\caption{Scanning photovoltage maps for excitations at a) 514nm, b) 633nm, c) 785nm and perpendicular polarization (TM). The grating coupler dimensions are not to scale in the vertical direction. d) Photovoltage at 785nm as a function of polarization. 0$^\circ$ denotes TM polarization. }
\label{graph:grating_pv}
\end{figure*}

Fig.\ref{fig:biosensor} numerically tests this idea simulating four pairs of sensor arms, tuned to operate at 678nm according to Fig.\ref{fig:theory1}c: (i) $d_1$=410nm, $d_2$=750nm, (ii) $d_1$=775nm, $d_2$=1075nm, (iii) $d_1$=1070nm, $d_2$=1400nm and (iv) $d_1$=1435nm, $d_2$=1730nm. In cases (i)-(iv) the left arm alternates from being on a positive to a negative slope of the interference diagram of Fig.\ref{fig:theory1}c. Thus, the photovoltage in the presence of the analyte is also expected to alternate. For simplicity, the analyte is assumed to be a thin film deposited on the sensor arms and to have an $n$=1.55, an average value for dry protein films\cite{voros04}. Fig.\ref{fig:biosensor}b plots the asymmetric absorption as a function of analyte thickness. We obtain a linear response, with slope increasing the longer the sensor arms. In Fig.\ref{fig:biosensor}c the absolute value of the slope is plotted as a function of the small sensor arm length $d_1$, and a good linear fit is obtained. Tuning the arm dimensions thus provides an additional tool for controlling and tuning the device's performance and sensitivity. At long arm lengths, SPP losses will limit the sensitivity, but they could be overcome, e.g. by increasing the metal thickness and/or reducing the substrate index to limit SPP coupling to the substrate.

Having demonstrated the design versatility of grating-coupled GPDs, we now turn to the experimental validation of our predicted SPP enhanced SLG absorption and photodetection. The general architecture of our experimental devices with plasmonic grating coupler is shown in Fig.\ref{graph:graphene_grating}a. SLG is contacted with metallic source and drain electrodes. In order to quantify the enhancement relative to the no-grating case, we fabricate the grating coupler on top of one of the contacts and leave the other contact flat. This is the simplest way to break the contacts symmetry and allows the generation of net non-zero photoresponse, even with both contacts illuminated.

Our devices are fabricated as follows. Graphene is produced by mechanical exfoliation of graphite onto an Si+300nm SiO$_2$\cite{Novoselov2004,Novoselov2005} and characterized by optical microscopy\cite{Casiraghi2007} and Raman spectroscopy\cite{Ferrari2006,FerrariNatNano2013}. Subsequently, the source and drain contacts are prepared by e-beam lithography and a base metallization layer is deposited by thermal evaporation of 4nm Cr and 50nm Au, employing a lift-off step. The 620nm period grating is then defined in a further e-beam lithography step by performing a second thermal evaporation of 50nm Au followed by lift-off. Fig.\ref{graph:graphene_grating}b shows a scanning electron micrograph (SEM) of the device. A slight asymmetry between the ridges and grooves is detected due to overexposure during e-beam lithography, but this does not change the spectral characteristics of the grating, solely determined by its period. Afterwards, the samples are bonded into a chip carrier for electrical and optical characterization.

We perform wavelength dependent photovoltage mapping to determine the spatial pattern of the devices' photoresponse. Fig.\ref{graph:grating_pv} plots the photovoltage maps at different incident wavelengths for polarization perpendicular to the grating (TM-polarization, 100x ultralong working distance objective, numerical aperture NA=0.6). Fig.\ref{graph:grating_pv} also shows the structured grating contact and the flat contact without perturbations. At 514nm the photoresponse occurs predominantly at the contacts' edges and is of similar magnitude, but opposite polarity, to that of a standard MGM PD\cite{Lee2008,Mueller2009}. At 633nm, Fig.\ref{graph:grating_pv}b indicates that the influence of the grating starts emerging. 2-3$\mu$m away from the edge of the patterned contact, a photoresponse is visible, even though no junction is present. The effect is much more pronounced at 785nm, where the entire structured contact becomes photosensitive, Fig.\ref{graph:grating_pv}c, and the photoresponse is enhanced$\sim$400\% compared to the flat contact edge. Furthermore, the responsivity of the device is polarization dependent, as shown in Fig.\ref{graph:grating_pv}d. The strongest photoresponse occurs for perpendicular polarization (0$^\circ$,TM-polarized light).
\begin{figure}
\centerline{\includegraphics[width=90mm]{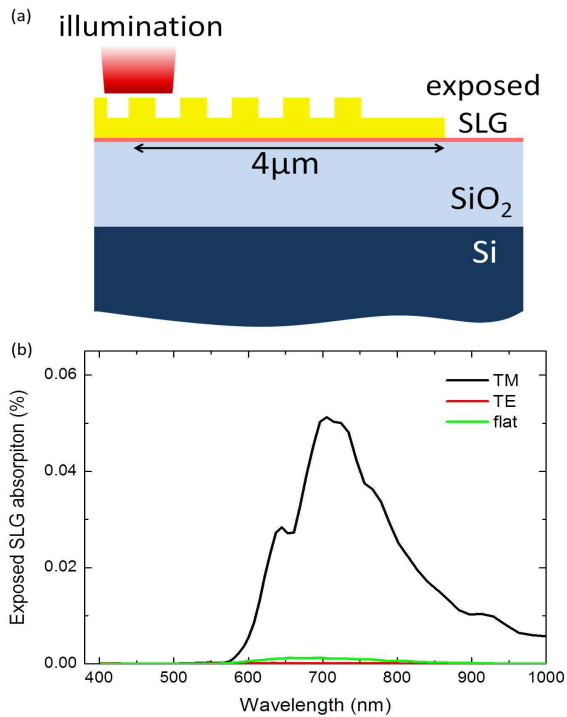}}
\caption{Calculated absorption in the SLG exposed part when a 1$\mu$m spot size illuminates the grating 4$\mu$m away from the metal/SLG junction. (a) Schematic of the simulation system. 300nm SiO$_2$ on Si is assumed for closer matching to the experiments. (b) Absorption in SLG. Two polarizations and a flat contact case are considered.}
\label{graph:theory_focused}
\end{figure}

To have a direct comparison with the experiments, we perform calculations with a focused beam illumination (1$\mu$m width) on top of the grating,$\sim$4$\mu$m away from the contact edge, as depicted in Fig.\ref{graph:theory_focused}a. The termination step length is taken as 1250nm (estimated from the SEM image in Fig.\ref{graph:graphene_grating}) and we also include the Si substrate with 300nm SiO$_2$. Fig.\ref{graph:theory_focused}b plots the absorption in the exposed SLG 4$\mu$m away from the illumination spot for TM polarization, TE polarization and a flat contact. Both the theoretical spectral and polarization responses are in excellent agreement with experiments, and verify the strong responsivity above 700nm. This contrasts with the normally incident plane wave illumination case, below 700nm (see Fig.\ref{fig:theory1}b). This is understood by considering Eq.\ref{eq:spp2} and Fig.\ref{fig:theory2}f. For a normally incident plane wave and focused illumination, $k_{||}$ is zero and nonzero, respectively. Since the latter case is less restrictive for SPP excitation, it results into a wider SPP spectrum at both Au/air and Au/SiO$_2$ interfaces, thus into a wider responsivity compared to the plane wave case (see Methods for details)

In conclusion, we demonstrated the coupling to graphene of surface plasmon polaritons excited in a metallic plasmonic grating and its exploitation in graphene-based photodetection with enhanced responsivity and polarization selectivity. Depending on its dimensions, highly tunable spectral selectivity below 50nm bandwidth can be achieved. Further, the symmetry of the photodetector can be broken making it operable under full illumination, despite identical metal source and drain contacts. The underlying mechanism involves the coupling of light into SPPs on the patterned contact, and their propagation to the exposed SLG area. For uniform coherent illumination, these SPPs can further interfere with the waves directly incident on the exposed SLG, offering a novel tuning capability where the light flux can be attracted or repelled from the contact edge by design. The whole contact thus becomes a highly tunable polarization- and spectral-selective photosensitive area. SPPs and incident wave interference can potentially be employed for (bio-) sensing by tailoring the grating dimensions. This may allow a novel plasmonic sensing architecture with high sensitivity and small footprint with direct electrical readout and without complicated optics.\\

We acknowledge funding from EU Graphene Flagship (no. 604391), ERC Grant Hetero2d, EPSRC Grants EP/K01711X/1, EP/K017144/1, EU grant GENIUS, a Royal Society Wolfson Research Merit Award.
\section{Methods}
\subsection{FDTD simulations}
\begin{figure}
\centerline{\includegraphics[width=90mm]{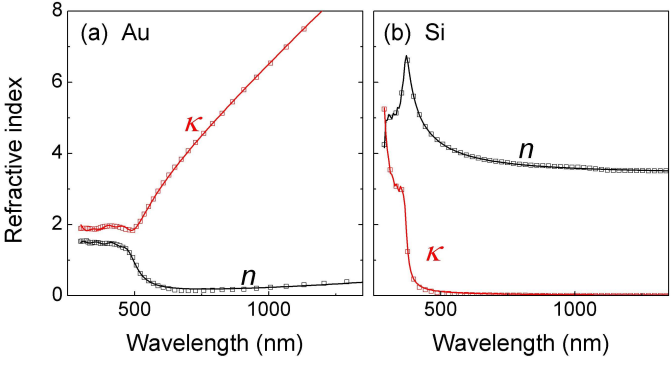}}
\caption{Refractive indexes used in the calculations for a) Au, b) Si. Symbols are experimental data from Refs.\cite{johnsonchristy2,palik}; lines are the corresponding Drude-Lorentz fits.} \label{fig:theoryS1}
\end{figure}

In our FDTD simulations Maxwell's equations are time-integrated on a computational grid, with a Drude-Lorentz model assumed for the dielectric function\cite{wooten}:
\begin{equation}
\epsilon(\omega)=\epsilon_{\infty}-\frac{\omega_p^2}{\omega^2+i\omega\gamma}+\sum_{j=1}^N\frac{\Delta\epsilon_j\Omega_j^2}{\Omega_j^2-\omega^2-i\omega\Gamma_j}
\end{equation}
where the first term is the Drude free-electron contribution and the second contains Lorentz oscillators corresponding to interband transitions. $\omega_p$ and $1/\gamma$ are the free electron plasma frequency and relaxation time, $\Omega_j$, $\Delta\epsilon_j$, $\Gamma_m$ are transition frequency, oscillator strength and decay rate for the Lorentz terms. To accurately reproduce the experimental dielectric functions (Au from Ref.\cite{johnsonchristy2} and Si from Ref.\cite{palik}) we treat these as fit parameters. For Au we use $N$=4, and $\epsilon_\infty$=4.054, $\Delta\epsilon_j$=(0.43, 0.634, 0.755, 1.059), $\hbar\omega_p$=8.76eV, $\hbar\gamma$=0.068eV, $\hbar\Omega_j$=(2.67, 3.03, 3.54, 4.23)eV, and $\hbar\Gamma_j$=(0.458, 0.641, 0.892, 0.959)eV. For Si we use $N$=7 without any Drude term: $\epsilon_\infty$=1.89, $\Delta\epsilon_j$=(1.198, 0.963, 1.021, 1.164, 1.407, 2.259, 1.869), $\hbar\Omega_j$=(3.39, 3.51, 3.68, 3.86, 4.06, 4.25, 4.61)eV, and $\hbar\Gamma_j$=(0.188, 0.203, 0.239, 0.269, 0.283, 0.265, 0.0)eV. Fig.\ref{fig:theoryS1} plots our model dielectric functions along with the experimental ones, showing an excellent agreement.

The computational cell comprises a 2nm cubic grid. Perfectly matched layer (PML) boundary conditions\cite{pml} are applied at the edges of the cell in the vertical direction. In the lateral direction, we use periodic boundary conditions for the uniform illumination case, and PML for the focused.
\subsection{SLG index of refraction}
\begin{figure}
\centerline{\includegraphics[width=90mm]{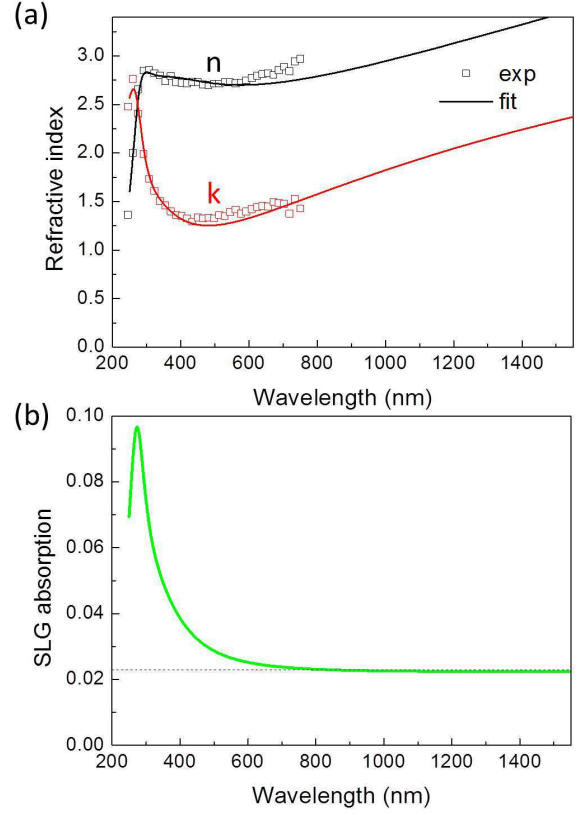}}
\caption{(a) SLG refractive index used in our calculations. Symbols are ellipsometric data from Ref.\cite{kravets10}. (b) SLG absorption corresponding to the applied model.}
\label{fig:theoryS2}
\end{figure}
\begin{figure*}
\centerline{\includegraphics[width=150mm]{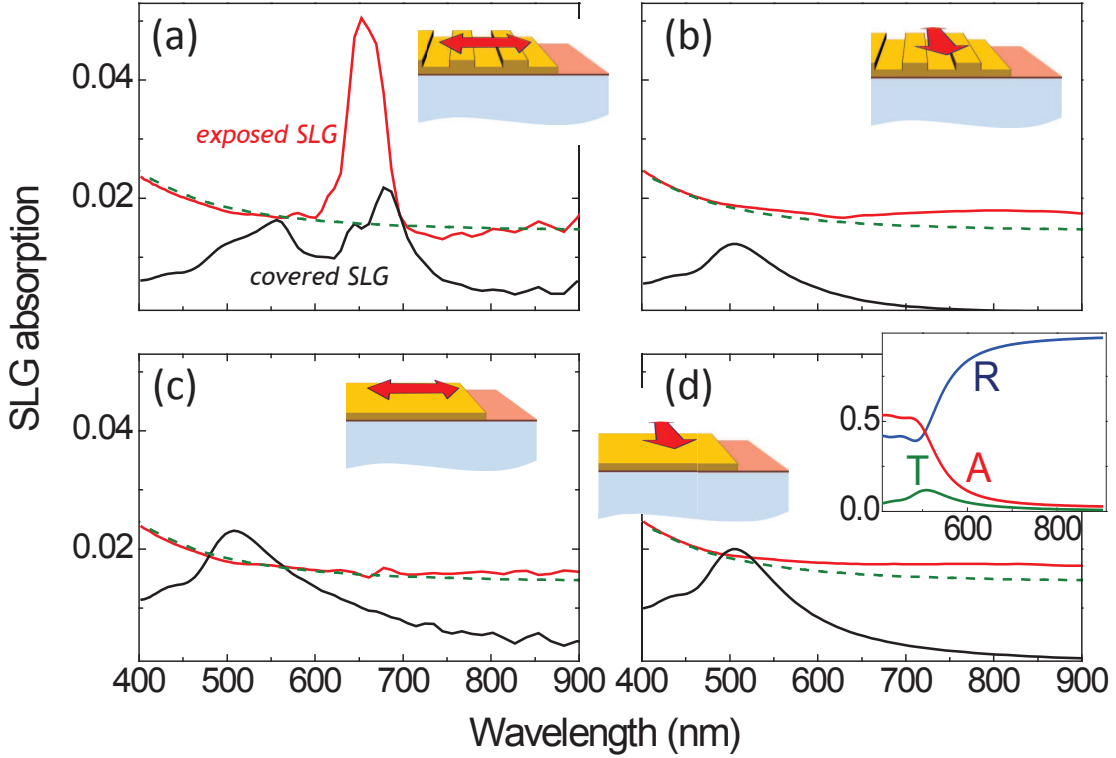}}
\caption{Absorption in the 1$\mu$m exposed part of SLG (red line) and in the covered SLG under the Au contact (black line). The green dotted line is the absorption in 1$\mu$m of SLG on top of SiO$_2$. All curves are normalized to the light flux incident in 1$\mu$m. (a) TM-polarized, (b) TE-polarized. (c,d) The two polarizations for unpatterned contacts. In cases (b,d) a peak at$\sim$500nm is observed. This is related to a transparency window in Au, as shown in the inset where the reflection, absorption and transmission through a 50nm thick Au film suspended in air are plotted.}
\label{fig:theoryS3}
\end{figure*}
The index of refraction of SLG was measured by ellipsometry for wavelengths up to 750nm in Ref.\cite{kravets10}. To extrapolate to longer wavelengths, we assume undoped SLG and use the universal optical conductance $\sigma=\mathrm{e}^2/4\hbar$, to get the dielectric function:
\begin{equation}
\epsilon=\epsilon_{\infty}+i\frac{4\pi\sigma}{\omega d_{SLG}}=\epsilon_{\infty}+i\frac{\alpha \lambda}{2d_{SLG}}
\label{eq:slg_index}
\end{equation}
where $d_{SLG}$=0.335nm is SLG's thickness and $\alpha=\mathrm{e}^2/\hbar c$ is the fine structure constant. In the thin film limit, Eq.\ref{eq:slg_index} yields:
\begin{equation}
\label{eq:slg_abs}
A_{SLG}^0 \cong \mathrm{Im}(\epsilon)\frac{2\pi d_{SLG}}{\lambda} = \pi\alpha = 2.3\%
\end{equation}
A value of $\epsilon_{\infty}=5.7$ ensures that Eq.\ref{eq:slg_index} matches the ellipsometric experimental data of Ref.\cite{kravets10} at smaller wavelengths. We fit a Drude-Lorentz model to both the ellipsometric data at small wavelengths and Eq.\ref{eq:slg_index} at longer wavelengths, Fig.\ref{fig:theoryS2}a. In Fig.\ref{fig:theoryS2}b the resulting SLG absorption is plotted. The Drude-Lorentz model for SLG uses $N$=4 with $\epsilon_\infty$=2.148, $\Delta\epsilon_j$=(64.8, 2.92, 1.69), $\hbar\omega_p$=1.34eV, $\hbar\gamma$=0.7eV, $\hbar\Omega_j$=(1.0, 4.0, 4.56)eV, and $\hbar\Gamma_j$=(5.41, 2.77, 1.0)eV. Within FDTD's 2nm grid, SLG is treated as an effective 2nm thick slab, thus its dielectric function is scaled to reproduce the correct absorption and reflection properties according to $\epsilon \rightarrow 1+ (\epsilon-1)d_{SLG}/2$, i.e. $\epsilon_\infty\rightarrow$1.192 and $\Delta\epsilon_j\rightarrow$(10.85, 0.488, 0.283).
\subsection{SLG absorption in the thin film limit}
In a three-layer system, the normal incidence Fresnel equations for reflection and transmission amplitudes are\cite{anders,hecht}:
\begin{eqnarray}
r&=&r_{12}+\frac{t_{12} r_{23} t_{21} \mathrm{e}^{2i\phi}}{1-r_{12}r_{23} \mathrm{e}^{2i\phi}}\\
t&=&\frac{t_{12}t_{23}\mathrm{e}^{i\phi}}{1-r_{12}r_{23} \mathrm{e}^{2i\phi}}
\end{eqnarray}
where $r_{ij}=(n_i-n_j)/(n_i+n_j)$, $t_{ij}=2n_i/(n_i+n_j)$ and $\phi =n_2 \omega d_2 /c$, with $d_2$ the film thickness. Here we assume the incoming medium $n_1\equiv n_{air}=1$, the film $n_2\equiv n_{SLG}$ and the semi-infinite substrate $n_3\equiv n_{SiO_2}$. Reflection and transmission coefficients are $R=|r|^2$ and $T=n_3|t|^2$. In the thin film limit, using the SLG dielectric function given by Eq.\ref{eq:slg_index}, we get:
\begin{eqnarray}
R&\cong &\left( \frac{n_1-n_3}{n_1+n_3}\right)^2 \left(1- \frac{4\beta n_1}{n_1^2-n_3^2} +\left|\frac{\beta(\epsilon-n_1n_3)}{\Im{\epsilon}(n_3-n_1)}\right|^2\right) \label{eq:refl}\\
T&\cong &\frac{4n_1 n_3}{(n_1+n_3)^2} \left(1- \frac{2\beta}{n_1+n_3}\right)
\end{eqnarray}
where $\beta=2\pi d Im\left\{\epsilon\right\}/\lambda=N\pi\alpha$, $\epsilon$ is the film dielectric constant, $\alpha=\mathrm{e}^2/\hbar c$ is the fine structure constant, and $d=Nd_{SLG}$ is the N-layer graphene (NLG) thickness. This results into the absorption:
\begin{equation}
A =1-R-T \cong N\pi \alpha \frac{4n_1}{(n_1+n_3)^2}
\end{equation}
while for a NLG suspended inside a uniform medium (i.e. $n_1=n_3=n$) we find $A=N\pi \alpha/n=2.3/n\%$. These equations are valid for N<10\cite{Casiraghi2007,Kuzmenko2008,Nair2008}. For $N>10$ the thin film limit breaks down and the optical paths inside the film must be taken into account.

Regarding reflection, this is typically dominated by the substrate since the difference in refractive index is usually largest between air and the substrate. However, in the $n_1=n_3=n$ case the third term in the right hand side of Eq.\ref{eq:refl} is the only nonzero term, resulting into NLG reflection: $R\cong N^2\pi^2 d_{SLG}^2|\epsilon-n^2|^2/(n^2\lambda^2)$. This is very small for SLG ($R\cong0.02\%$ at 600nm for suspended SLG\cite{Nair2008}) but it increases quadratically with N.
\begin{figure}
\centerline{\includegraphics[width=90mm]{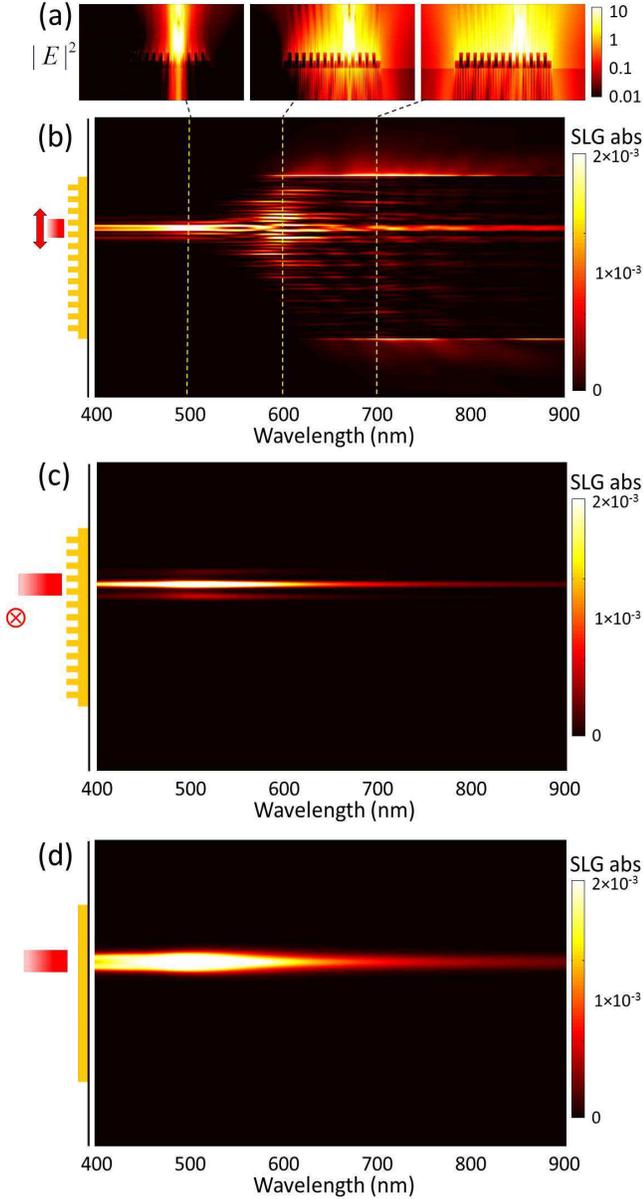}}
\caption{(a) Field intensity distribution for a 1$\mu$m spot illumination at normal incidence (TM), for 500, 600, 700nm. The color-coding is in logarithmic scale. (b) SLG absorption distribution (vertical axis) as a function of illumination wavelength (horizontal axis). The three wavelengths studied in (a) correspond to the three characteristic cases: absorption under the illumination spot, extended absorption under the contact, and absorption in the exposed SLG far from the illumination spot. (c) the same as in (b) but for TE polarization. (d) Same as above for unpatterned contacts (both polarizations yield identical results). Absorptions are normalized to the incident light flux.}
\label{fig:theoryS4}
\end{figure}
\subsection{Coherent Uniform illumination}
\begin{figure}
\centerline{\includegraphics[width=90mm]{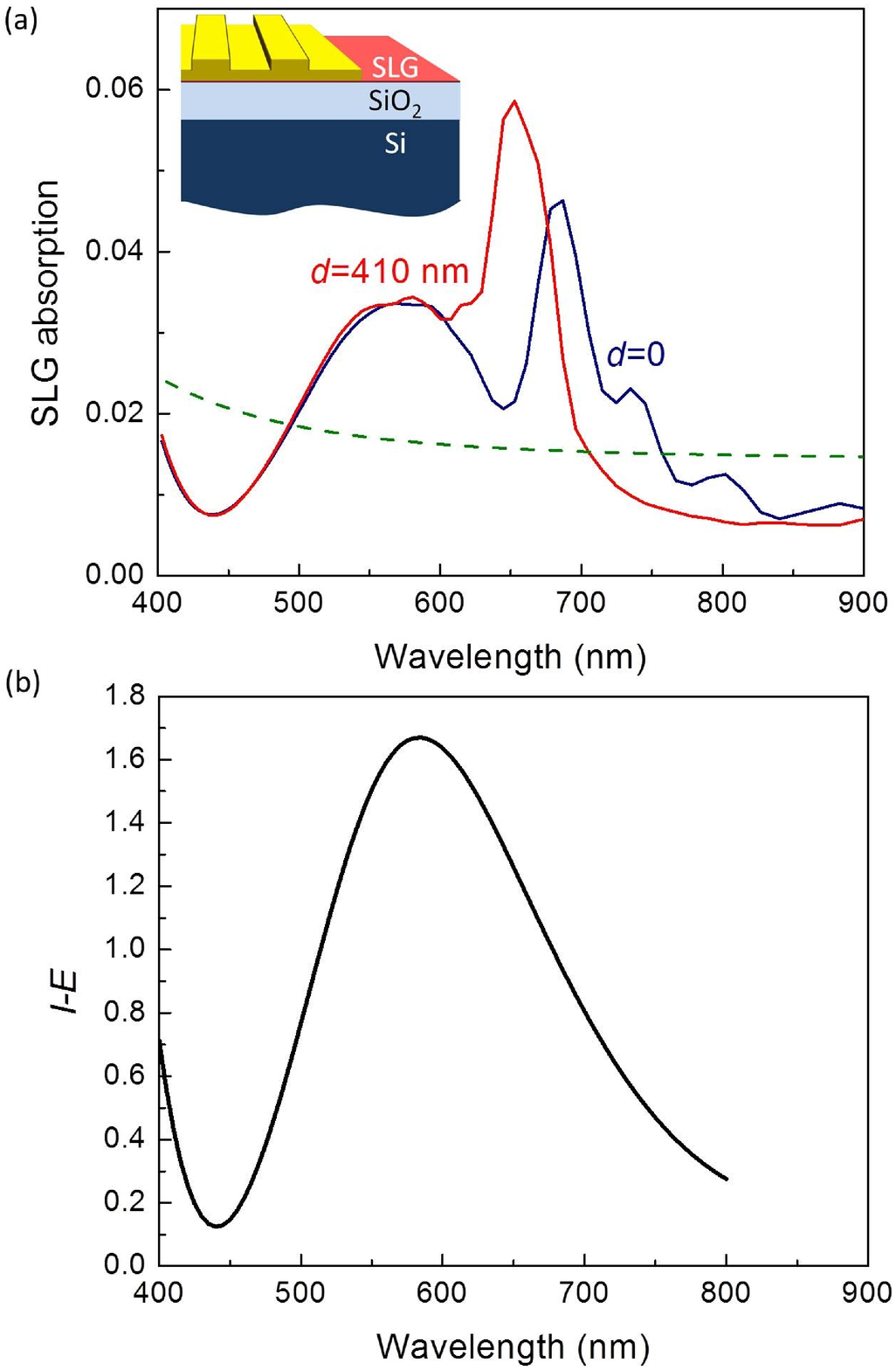}}
\caption{(a) Absorption in the exposed SLG for the system studied in Fig.1 with a SiO$_2$(300nm)/Si substrate. (b) Absorption interference enhancement without Au contacts. The modulation at$\sim$570nm is due to I-E.}
\label{fig:theoryS5}
\end{figure}
\begin{figure*}
\centerline{\includegraphics[width=165mm]{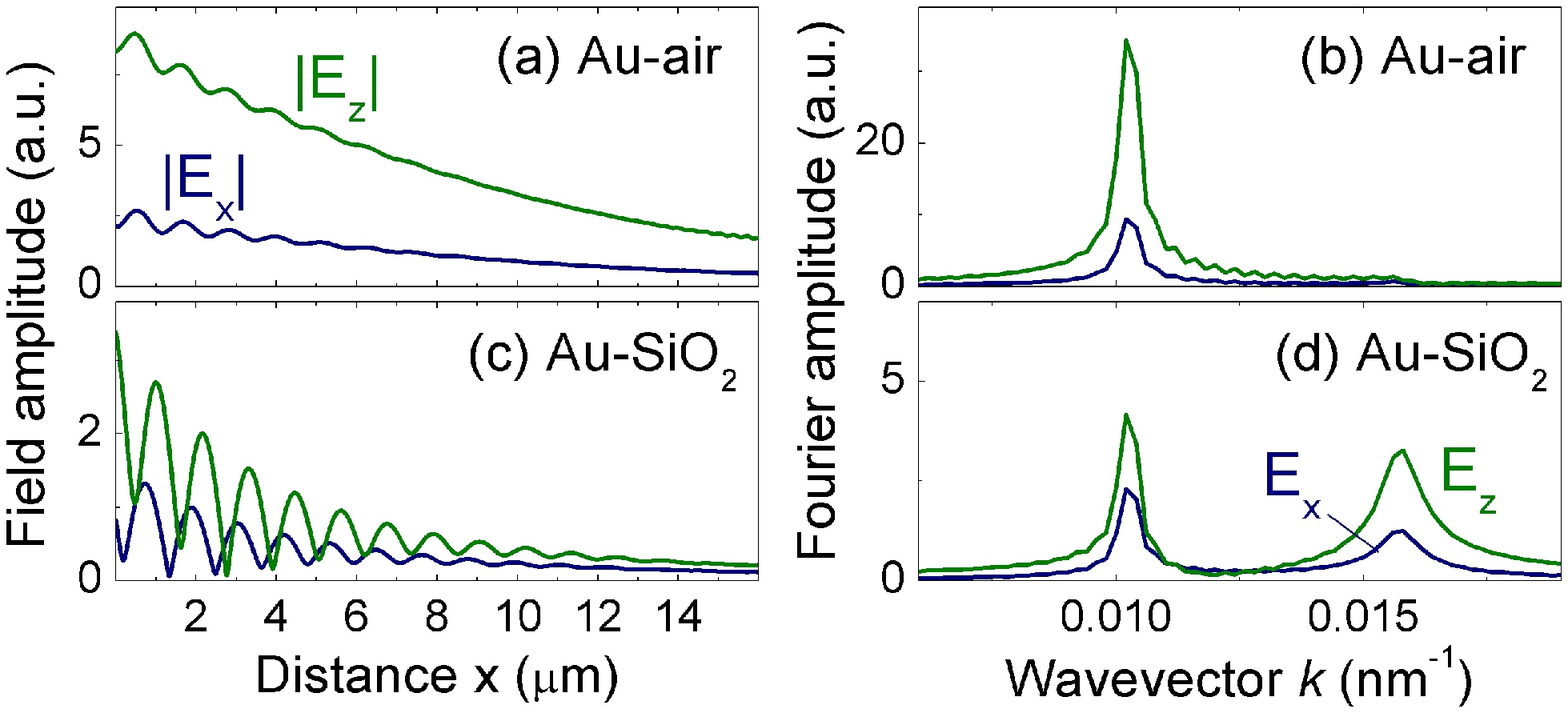}}
\caption{System of Fig.3 in with a SiO$_2$~(300~nm)/Si substrate. Besides a small modulation in the Fourier amplitudes, the SPP characteristics remain the same as for the semi-infinite SiO$_2$ substrate. Notable is also a deeper "beat" modulation of the field intensity at the Au-SiO$_2$ interface, because of reflection of the "leaked" SPP at the SiO$_2$/Si interface.}
\label{fig:theoryS6}
\end{figure*}
Fig.\ref{fig:theoryS3} plots the response of our system considering the absorption in both exposed and covered SLG for both polarizations and patterned and unpatterned contacts. For simplicity we only consider the $d$=410nm case with 13 ridge periods. We find that in all cases, other than TM-excitation on the patterned contact, the absorption in the exposed SLG is similar to that of a SLG on top of SiO$_2$ in the absence of the contact (green dotted line). Interestingly, there is some absorption in the covered SLG as well. In the TM-excitation of the patterned contact we obtain a similar modulation with wavelength as that in the exposed SLG, albeit of significantly smaller magnitude. We note that, in order to directly compare with the exposed SLG absorption, we still normalize to the power illuminating the exposed SLG area. Had we normalized to the flux illuminating the whole contact, then the covered SLG absorption would appear with a much smaller magnitude than in Fig.\ref{fig:theoryS3}. We also obtain absorption in the other cases as well, peaked at$\sim$500nm. This, however, is irrelevant to the grating and SPPs, as it is there that Au becomes most transparent. Au is strongly absorptive at high energies because of the onset of interband transitions from its d-electrons\cite{johnsonchristy2} while it is strongly reflective at small energies because of its conduction electrons\cite{johnsonchristy2}. The inset in Fig.\ref{fig:theoryS3}d plots the reflection, transmission and absorption coefficients through a 50nm thick Au film in air, displaying Au's transparency window at$\sim$500nm.
\subsection{Focused illumination}
We explore the case of focused illumination on the grating, which more closely resembles the experiments. To better facilitate the simulation and avoid having laterally scattered light re-enter the computational cell, we remove the lateral periodic boundary conditions and adopt PML boundary conditions\cite{pml}, so that any light scattered towards the sides of the computational cell permanently exits the calculations. We also consider a much larger exposed SLG area for better visualization. We adopt a TM polarized 1$\mu$m-wide plane source illuminating the 13-period $d$=410nm grating at a non-symmetric position, as depicted in Fig.\ref{fig:theoryS4}. The frequency domain electric-field intensity profiles for $\lambda$=500, 600 and 700nm are shown in logarithmic scale in Fig.\ref{fig:theoryS4}a. No scattering occurs at 500nm, while the most intense is seen at 700nm.

The system's full response is shown in Fig.\ref{fig:theoryS4}b, where we plot the SLG absorption throughout the length of the structure (vertical axis) for different wavelengths (horizontal axis). The illumination source is again a TM-polarized 1$\mu$m wide spot as shown in the inset schematic. Absorption is normalized to the peak incoming flux per unit area. Three distinct regions emerge: up to 550nm, absorption only occurs in the covered SLG directly underneath the illumination source. For 550nm$ <\lambda< $650nm, increased absorption is found in an extended area (several microns away from the illumination spot) in the covered SLG, pointing towards light diffraction into SPPs in the Au-SiO$_2$ interface. Above 650nm there is strong absorption in the exposed SLG beyond the grating. This points towards light diffraction into SPPs in the Au-air interface, which propagates and reaches the exposed SLG at the grating's edge. In Fig.\ref{fig:theoryS4}c we plot the same absorption map for TE illumination of a patterned contact, while in Fig.\ref{fig:theoryS4}d for illumination of an un-patterned contact. In both these cases no absorption is found, except directly underneath the illumination spot. It is thus clear that SPP-mediated effects are dominating the response for TM-polarized illumination.
\subsection{SiO$_2$/Si substrate effects}
The SiO$_2$ (300nm)/Si substrate has two effects. First, it provides some interference-based enhancement in the SLG absorption\cite{wang2008,son2009,gao2009,cinzPRB}. Fig.\ref{fig:theoryS5}a plots the exposed SLG absorption for the system described in Fig.1 on top of a 300nm SiO$_2$/Si substrate. The response is similar to the semi-infinite SiO$_2$ case, except for an overall modulation due to the interference effects in the SiO$_2$ dielectric spacer. The net interference-enhancement (I-E) effect on absorption (i.e. without the patterned contact) is plotted in Fig.\ref{fig:theoryS5}b. The dielectric spacer can thus be used as an additional degree of freedom (i.e. the spacer's index and thickness) in optimizing the system's response. We also simulated the asymmetric contact layouts studied in Fig.1d on top of 300nm SiO$_2$/Si and found that they produce a very similar response irrespective of the finite dielectric spacer. This is expected, since I-E partially cancels out when considering asymmetric absorption.

The second effect of the 300nm SiO$_2$/Si substrate is that the leaked SPP of the Au-air interface will be reflected back from the Si substrate. Fig.\ref{fig:theoryS6} plots the field intensities along the two Au interfaces as well as the corresponding Fourier transform amplitudes (see Fig.3b-e for the 300nm SiO$_2$/Si case). We note a deeper "beat" modulation of the fields on the Au-SiO$_2$ interface, due to the back reflected fields of the "leaked" Au-air SPP. This is also apparent in the Fourier amplitude. However, no frequency shifts are observed in the latter. We thus conclude that, other than some small amplitude modulations, the SPP structure is largely unaffected by the SiO$_2$ being 300nm thick or semi-infinite.

\end{document}